\documentclass[aps,pre,noshowpacs,twocolumn,groupedaddress]{revtex4}

\usepackage{hyperref}
\usepackage{amsmath}
\usepackage{amsfonts}
\usepackage{amssymb}
\usepackage{graphicx}
\usepackage{empheq}
\usepackage{diagbox}

\usepackage{tikz,xcolor,hyperref}

\definecolor{linkcolour}{HTML}{000066}	%light purple link for the email
\hypersetup{colorlinks,breaklinks,
	urlcolor=linkcolour, 
	linkcolor=linkcolour,
	citecolor=linkcolour}

\definecolor{lime}{HTML}{A6CE39}
\DeclareRobustCommand{\orcidicon}{
	\begin{tikzpicture}
		\draw[lime, fill=lime] (0,0) 
		circle [radius=0.16] 
		node[white] {{\fontfamily{qag}\selectfont \tiny ID}};
		\draw[white, fill=white] (-0.0625,0.095) 
		circle [radius=0.007];
	\end{tikzpicture}
	\hspace{-2mm}
}

\newcommand{\orcidauthorNB}{\href{https://orcid.org/0000-0002-1554-3820}{\orcidicon}} % Niall Byrnes
\newcommand{\orcidauthorMRF}{\href{https://orcid.org/0000-0001-5864-9636}{\orcidicon}} % Matthew Foreman
\newcommand{\orcidauthorGG}{\href{https://orcid.org/0000-0002-7342-4790}{\orcidicon}} % Gary Greaves

\hypersetup{colorlinks=true, urlcolor=blue, linkcolor=blue, citecolor=blue,plainpages=false}

\newcommand{\txtpow}[1]{{\mbox{\scriptsize{#1}}}}

\begin{document}
	\title{Bootstrapping cascaded random matrix models: correlations in permutations of matrix products }

        \author{Niall Byrnes\orcidauthorNB$^1$, Gary R. W. Greaves\orcidauthorGG$^2$ and Matthew R. Foreman\orcidauthorMRF$^{1,3}$}
	\email[]{matthew.foreman@ntu.edu.sg}
	\affiliation{
		$^1$School of Electrical and Electronic Engineering, Nanyang Technological University, 50 Nanyang Avenue, Singapore 639798 \\
		$^2$School of Physical and Mathematical Sciences, Nanyang Technological University,
		21 Nanyang Link, Singapore 637371, Singapore \\
		$^3$Institute for Digital Molecular Analytics and Science, 59 Nanyang Drive, Singapore 636921
	}
	\date{\today}

	\begin{abstract}
    	Random matrix theory is a useful tool in the study of the physics of multiple scattering systems, often striking a balance between computation speed and physical rigour. Propagation of waves through thick disordered media, as arises in for example optical scattering or electron transport, typically necessitates cascading of multiple random matrices drawn from an underlying ensemble for thin media, greatly increasing computational burden. Here we propose a dual pool based bootstrapping approach to speed up statistical studies of scattering in thick random media. We examine how potential matrix reuse in a pool based approach can impact statistical estimates of population averages. Specifically, we discuss how both bias and additional variance in the sample mean estimator are introduced through bootstrapping. In the diffusive scattering regime, the extra estimator variance is shown to originate from samples in which cascaded transfer matrices are permuted matrix products. Through analysis of the combinatorics and cycle structure of permutations we quantify the resulting correlations. Proofs of several analytic formulae enumerating the frequency with which correlations of different strengths occur are derived. Extension to the ballistic regime is briefly considered. 
	\end{abstract}
	
	\maketitle

	\section{Introduction}
	Computational modelling of wave scattering in random disordered media is a difficult problem that has been researched in earnest in recent decades. Full finite element based solution of the relevant wave equation offers the greatest rigour, but consequently suffers from limited simulation volumes and large computational cost \cite{Pamel2017}. Alternatively, more approximate methods, such as diffusion, Monte Carlo or Green's function based approaches balance computation time with physical rigour to varying degrees \cite{Rossum1999,Meglinski2005,RamellaRoman2005,Markel2019}. Random matrix approaches, in particular, typically sacrifice system-specific details in favour of significantly reduced computation times, allowing statistical properties, such as mean transmission, spectral distributions and phase transitions, to be numerically investigated \cite{Beenakker1997,Akemann2015}. Knowledge of such general features of random scattering media has afforded great physical insights \cite{Aiello2005a,Choi2011a,Hanada2018,Byrnes2020,Garratt2021} and enabled development of a number of useful techniques, for example, for imaging through scattering media and wavefront control \cite{Aubry2009,Rotter2017,Hsu2017}. 
	
	Early random matrix models such as the circular and Gaussian ensembles proposed by Wigner and Dyson \cite{Wigner1955,Dyson1962,Dyson1962a}, were relatively simplistic and were based on the assumption of isotropic scattering which often poorly approximates reality. More sophisticated random matrix models have however been developed in an attempt to capture additional symmetries, constraints or system specific properties, whilst preserving numerical speed \cite{Andreev1994,Akemann2015,Byrnes2022a}. For example, filtered matrix ensembles can more accurately simulate transmission in systems where the input and output measurement channels are a finite subset of all potentially available channels \cite{Goetschy2013}, whereas Euclidean matrix ensembles can describe random Green's matrices relevant to propagation of waves in collections of point-like scattering centers \cite{Goetschy2011,Goetschy2013a,Berk2021b}. Wishart and Jacobi ensembles meanwhile are also classic ensembles useful, for instance, in the description of chaotic cavities \cite{Vivo2008, Mehta2004}. Cascaded, or coupled, random matrix models have also seen significant study in the literature finding applications in e.g., electronic transport in wires, quantum chaos, wireless communications, study of spin glasses \cite{Beenakker1997,Rotter2017,Tulino2004,Crisanti} and more. Such models describe the linear scattering properties of a system, as described by its (random) transfer matrix $\mathbb{T}$, through the correctly-ordered product of individual transfer matrices, $\mathbb{T}^{\delta}$, (drawn from an appropriate underlying matrix ensemble) of constituent scattering sites or system components. Cascaded models can, for example, describe universal conductance fluctuations \cite{Weidenmuller1990} and naturally introduce a notion of length to the system such that variations in total transmission and reflection with system size can be modelled (in contrast to more traditional matrix ensembles). Of particular significance is the Dorokhov-Mello-Pereyra-Kumar (DMPK) cascaded matrix model \cite{Dorokhov1982,Mello1988b} and its higher dimensional and electromagnetic generalisations \cite{Muttalib2002,Douglas2014,Byrnes2022b}, which recognise that a thick scattering system can be considered as a chain of thinner, weakly scattering, media. Each successive matrix in the associated product acts as a perturbation, enabling description of various wave transport regimes in a scattering medium (e.g., ballistic, diffusive, localized), through cascades of different lengths. In the so-called ballistic scattering regime, the average number of scattering events of a wave propagating in the medium is $\ll 1$, whereas in the diffusive regime, corresponding to thicker media, transmitted waves undergo many scattering events.  
     
Although allowing a more physically accurate description of a large variety of scattering systems, cascaded random matrix techniques can suffer from reduced speed since calculation of the transfer matrix for a single realisation of a thick medium can necessitate generation of a large number of thin slab matrices. Moreover, statistical convergence is slower for large thicknesses due to the greater variances typically involved thereby requiring more realisations to be simulated and hence entailing longer computation times. In this work, our aim is to investigate computational gains that can be made in cascaded random matrix models through a bootstrapping approach employing pools of pre-calculated matrices. Moreover, we seek to assess the extent to which use of a pool based bootstrapping approach can degrade statistical estimates of mean scattering properties of thick scattering media. In Section~\ref{sec:dualpool}, we first introduce and discuss a dual pool bootstrapping method, which can reduce both the number of matrix generations and matrix products required for statistical studies of system properties. In Section~\ref{sec:biasvar}, we investigate estimator bias and variance when mean properties are derived from transfer matrices formed from products of matrices sampled from a matrix pool $\mathcal{P}$. We show that bootstrapping can introduce additional estimator fluctuations, the prevalence of which are enumerated in Section~\ref{sec:enum} (supporting proofs based on combinatorics of permutations are presented in Appendix~\ref{appendix}). Finally, in Section~\ref{sec:generalisations}, we briefly discuss extending our findings to more complex scenarios. 

 \section{Dual pool bootstrapping of cascaded random matrices}\label{sec:dualpool}

 Computational gains in cascaded random matrix models can be made if a bootstrapped approach is adopted in which individual matrices $\mathbb{T}^\delta$ are resampled from an existing set instead of on-the-fly generation. This sample could, for example, represent a pool of pre-generated transfer matrices or experimental data. To illustrate this principle, consider simulating $N_r$ realisations of the transfer matrix of a medium of thickness $L$ through cascading $N_{\delta}$ slabs of thickness $\delta L = L/N_{\delta}$ using on-the-fly calculations.  As indicated in Table~\ref{tab:numcalcs} this would require $N_r N_{\delta}$ thin-slab matrices to be generated and calculation of $N_r (N_\delta -1)$ matrix products. Alternatively,  pre-generation of a single pool $\mathcal{P}_1$ of $N_1$ such matrices, from which $\mathbb{T}^{\delta}$ are resampled as needed, can greatly reduce the number of matrix generations required if $N_1 \ll N_r N_{\delta}$. With a single matrix pool, however, the required number of matrix products is unaffected (see Table~\ref{tab:numcalcs}). In scenarios where the number of relevant scattering modes can be very large, e.g., optical scattering, products of large transfer matrices can hence still limit total computation time and it is thus preferable to also reduce the number of matrix products. To this end, a dual pool approach can be adopted whereby the  pool $\mathcal{P}_1$ of $N_1$ thin-slab matrices is again pre-generated, however, additionally a second pool $\mathcal{P}_2$ of $N_2$ transfer matrices $\mathbb{T}^{\Delta}$ for sections of intermediate thickness $\Delta L$ is calculated. Each matrix in $\mathcal{P}_2$ is found by cascading $n_{\Delta} = \Delta L / \delta L$ thin-slab matrices which are themselves drawn from $\mathcal{P}_1$. The dual pool approach then allows the number of matrix products to be reduced to $N_2 n_{\Delta} + N_r (N_\Delta -1)$ where $N_\Delta = L/\Delta L$. It is also worthwhile to note that matrix pools can be stored and reused. By way of an example of potential gains, we note that in our earlier work \cite{Byrnes2022b} use of the dual pool approach allowed the number of matrix generations and products to be reduced by approximately three and two orders of magnitude respectively. Consequently, prohibitive computation times were overcome enabling statistical characterisation of a variety of polarisation scattering phenomena, such as light transmission and optical depolarisation, in multiply scattering media up to thicknesses of 30 mean free paths,  using $10^4$ realisations for each length. Physical parameters extracted from our simulations using the bootstrapped approach were found to be in  agreement with previously reported results.
	\begin{table}[t]
		\centering
		\begin{tabular}{ |c||c|c|}		
			\hline
			& Matrix generations & Matrix products\\
			\hline	\hline
			On-the-fly & $N_r N_\delta$&  $N_r (N_\delta -1)$\\
			Single pool &$N_1$ & $N_r (N_\delta -1)$\\
			Dual pool & $N_1$ & $N_2 n_{\Delta} + N_r (N_\Delta -1)$ \\
			\hline
		\end{tabular}
		\caption{Number of calculations required for on-the-fly, single pool or dual pool approaches.\label{tab:numcalcs}}
	\end{table}

	Resampling, or bootstrapping \cite{Efron1979,Efron1982,Efron1993}, approaches, such as the single or dual pool technique described, although enabling computational gains, do come at the potential cost of matrix reuse since, typically, matrices are sampled with replacement. In a single pool approach, a given transfer matrix $\mathbb{T}^\delta$ sampled from  $\mathcal{P}_1$ could in principle be used multiple times in calculation of $\mathbb{T}$. Similarly, for a dual pool approach a given $\mathbb{T}^\delta$ from $\mathcal{P}_1$ could be used several times when computing  single,  or different, $\mathbb{T}^{\Delta}$ matrices. Moreover, a $\mathbb{T}^{\Delta}$ drawn from $\mathcal{P}_2$ could be reused during generation of one or more instance of $\mathbb{T}$. Such matrix reuse is undesirable since it can introduce residual and unphysical biases and correlations between individual random matrices. It is therefore, important to establish the nature and quantify the magnitude of these detrimental effects, to assess the viability of use of the bootstrapped approach. This task forms the focus of the remainder of this work.

\section{Bootstrapped estimation of population averages}\label{sec:biasvar}

	We begin our analysis by first adopting a more general and convenient notation to more easily accommodate the different cases discussed above. Specifically, we consider an ensemble of random $M\times M$ transfer matrices $\mathbb{X}$ governed by the probability density function $p_{\mathbb{X}}(\mathbb{X})$. As an attempt to avoid notational clutter in what follows, we shall assume that transfer matrices are real, although extension of our results to complex ensembles is straightforward as will be discussed below. We also construct a pool $\mathcal{P}$ of $N_{X}$  transfer matrices $\mathbb{X}^{(j)}$, where we  label each matrix in the pool with the superscript $j \in [1,N_X]$. For on-the-fly calculations we can form a single realisation of a cascaded transfer matrix, $\mathbb{Z}$, by generating $N$ individual independent $\mathbb{X}$ matrices according to the true underlying ensemble probability distribution $p_{\mathbb{X}}(\mathbb{X})$. The $\mathbb{Z}$ matrices produced in this manner define a second ensemble governed by the probability density function $p_{\mathbb{Z}}(\mathbb{Z})$. Alternatively, when using a pool based approach we instead construct realisations of cascaded transfer matrices, $\mathbb{Z}^{\boldsymbol{\alpha}^{(l)}}$, by drawing matrices $\mathbb{X}^{(j)}$ from the pool $\mathcal{P}$ and evaluating the ordered matrix products 
	\begin{align}
		\mathbb{Z}^{\boldsymbol{\alpha}^{(l)}} = \prod_{p=1}^N \mathbb{X}^{(\alpha_p^{(l)})} = \mathbb{X}^{(\alpha_1^{(l)})}\mathbb{X}^{(\alpha_2^{(l)})}\cdots \mathbb{X}^{(\alpha_N^{(l)})},\label{eq:Z_def}
	\end{align}	
	where $\alpha_p^{(l)}$ is the $p$th member of a sequence $\boldsymbol{\alpha}^{(l)}$. The members of this sequence correspond to the labels of $N$ matrices drawn from the pool, such that $\boldsymbol{\alpha}^{(l)}$ defines the $l$th realisation of $\mathbb{Z}$. Note, the order of matrix products defined in Eq.~\eqref{eq:Z_def} will be assumed throughout this work. As an illustration to clarify the connection with our earlier discussion, we can make the associations $\mathbb{X} = \mathbb{T}^\delta$ and $\mathbb{Z}=\mathbb{T}^\Delta$, as corresponds to a dual pool approach. One possible sampling sequence  $\boldsymbol{\alpha}^{(1)} = (2,2,1,2,1)$, then means that the first realisation of $\mathbb{T}^\Delta = \mathbb{Z}^{\boldsymbol{\alpha}^{(1)}}$ is formed from sampling the first matrix in $\mathcal{P} = \mathcal{P}_1$ ($\mathbb{X}^{(1)}$)  twice and the second  matrix ($\mathbb{X}^{(2)}$) thrice, and taking the appropriately ordered matrix product.  Generation of multiple realisations of $\mathbb{Z}^{\boldsymbol{\alpha}^{(l)}}$ can then be collected to form $\mathcal{P}_2$.  Alternative associations include $(\mathbb{X},\mathbb{Z}) = (\mathbb{T}^{\delta},\mathbb{T})$ for the single pool approach or $(\mathbb{T}^{\Delta},\mathbb{T})$ for the dual pool approach.

	\subsection{Estimator bias}\label{sec:bias}

	Consider now the process of estimating some mean property, $F=\langle f(\mathbb{Z}) \rangle$, of the true matrix ensemble, where $\langle \ldots \rangle$ denotes the expectation with respect to $p_\mathbb{Z}(\mathbb{Z})$  and where $f$ is some arbitrary function. We may, for instance, wish to estimate mean transmission, channel capacity or medium depolarisation \cite{Nieuwenhuizen1995,Byrnes2020,Bicout1992}. In practice, to estimate $F$ we would consider the sample mean $\hat{F} = \bar{f} = {N_Z}^{-1}\sum_{l=1}^{N_Z} f(\mathbb{Z}^{\boldsymbol{\alpha}^{(l)}})$ where $N_Z$ is the number of realisations of $\mathbb{Z}$ that we choose to generate (or sample from a second pool) and where $\hat{F}$ denotes the estimator of $F$. The quality of $\hat{F}$ can be assessed through evaluation of the estimator bias 
	\begin{align}
		\mbox{bias}[\hat{F}]  = \langle \hat{F}  \rangle - \langle f\rangle.
	\end{align} 
	As an example, consider estimating the mean transfer matrix, $F = \langle\mathbb{Z}\rangle$, i.e., $f(\mathbb{Z}) = \mathbb{Z}$, using the sample mean,
	\begin{align}
		\bar{\mathbb{Z}} = \frac{1}{N_Z} \sum_{l=1}^{N_Z} \mathbb{Z}^{\boldsymbol{\alpha}^{(l)}} = \frac{1}{N_Z} \sum_{l=1}^{N_Z} \prod_{p=1}^N \mathbb{X}^{(\alpha_p^{(l)})}. \label{eq:Zbar}
	\end{align}
	Notably, if all indices in each sequence $\boldsymbol{\alpha}^{(l)}$ are distinct then taking the ensemble average of Eq.~\eqref{eq:Zbar} reduces to taking the product of the ensemble means of $\mathbb{X}$, i.e.,
	\begin{align}
		\langle \bar{\mathbb{Z}} \rangle = \frac{1}{N_Z} \sum_{l=1}^{N_Z} \prod_{p=1}^N \langle \mathbb{X}^{(\alpha_p^{(l)})} \rangle = \frac{1}{N_Z} \sum_{l=1}^{N_Z}  \langle \mathbb{X} \rangle^N = \langle \mathbb{Z} \rangle
	\end{align}
since each sample of $\mathbb{X}$ is drawn independently from the underlying ensemble and is identically distributed. In this scenario it is seen that the sample mean of the matrices $\mathbb{Z}^{\boldsymbol{\alpha}^{(l)}}$ formed from sampling the underlying pool $\mathcal{P}$ is an unbiased estimator of the true ensemble mean. It is simple to show this is true for more general functions $f(\mathbb{Z})$. When sampling matrices from the (finite) pool $\mathcal{P}$, however, there is a non-zero probability that the same matrix is drawn multiple times when sampling is performed with replacement, i.e., there may be repeated indices in $\boldsymbol{\alpha}^{(l)}$. Such matrix repetitions, in general, will destroy the unbiased nature of our estimator. Continuing our illustrative example, we can express the sample mean of the $(j,k)$th matrix element, $Z_{jk}$, in terms of explicit sums over the matrix elements of the constituent $\mathbb{X}^{(\alpha_p^{(l)})}$ matrices such that
\begin{align}
    \langle \bar{{Z}}_{jk} \rangle =  \frac{1}{N_Z} \sum_{l=1}^{N_Z}\sum_{u_1, \hdots, u_{N-1}} \bigg\langle \prod_{p=1}^N X_{u_{p-1}u_p}^{(\alpha_p^{(l)})} \bigg\rangle \label{eq:Zbarjkavg},
\end{align}
where the sums over $u_1,\ldots, u_{N-1}$ run from 1 to $M$ and for notational simplicity we let $u_0 =j, u_N=k$. The average of the products appearing in Eq.~\eqref{eq:Zbarjkavg} can be rewritten as the product of averages as before, however, factors for which $\alpha_p^{(l)}$ are equal must be grouped together. For instance, again considering the example whereby $\boldsymbol{\alpha}^{(1)} = (2,2,1,2,1)$, we can write
	\begin{align}
	\bigg\langle \prod_{p=1}^N X_{u_{p-1}u_p}^{(\alpha_p^{(1)})} \bigg\rangle =   \langle X_{u_{0}u_1}^{(2)} X_{u_{1}u_2}^{(2)}X_{u_{3}u_4}^{(2)}\rangle  
	\langle X_{u_{2}u_3}^{(1)}X_{u_{4}u_5}^{(1)} \rangle.
	 \label{eq:Zbarjkavg_example}
\end{align}
	In general, the average of a product, is not commensurate with the product of averages (e.g., $\langle X_{u_{2}u_3}^{(1)}X_{u_{4}u_5}^{(1)} \rangle \neq \langle X_{u_{2}u_3}^{(1)} \rangle\langle X_{u_{4}u_5}^{(1)} \rangle$), and can moreover be non-zero, particularly when the indices are the same (e.g., $u_2 = u_4$, $u_3 = u_5$). Consequently, it follows that when matrices are sampled from $\mathcal{P}$ with replacement the sample mean typically constitutes a biased estimator of the ensemble mean. Critically, the bias originates from the finite size of the pool from which we sample $\mathbb{X}^{(j)}$ and would not arise for on-the-fly calculations and is thus an artifact of the bootstrapped approach. A notable exception is when the $\mathbb{X}$ matrices belong to a zero mean Gaussian ensemble and each matrix product $\mathbb{Z}^{\boldsymbol{\alpha}}$ is formed from the product of an odd number of matrices ($N$ is odd). Since all odd moments of a zero mean Gaussian process are identically zero and at least one of the factors in Eq.~\eqref{eq:Zbarjkavg} will necessarily be the average of an odd number of terms, it follows that Eq.~\eqref{eq:Zbarjkavg} reduces to zero. More generally bias can be avoided when $\mathbb{X}^{(j)}$ are sampled from $\mathcal{P}$ without replacement when generating individual realisations of $\mathbb{Z}^{\boldsymbol{\alpha}}$ as will be assumed in the remainder of this work (although we note different $\mathbb{Z}^{\boldsymbol{\alpha}}$ matrices may crucially still have common $\mathbb{X}^{(j)}$ factors). Sampling without replacement, also more closely matches physical reality in the sense that the probability of any two (or more) sections of a thick random media being identical is infinitesimally small, such that if this possibility arises it is, again, purely an artifact of the sampling procedure that should be avoided.

	\subsection{Estimator variance}\label{sec:var}
	
	Having evaluated the bias of $\hat{F}$, we now consider the estimator variance $\mbox{var}[\hat{F}] = \langle \hat{F}^2\rangle - \langle \hat{F} \rangle^2$.  Using the definition of the sample mean it is simple to show 
	\begin{align}
		\mbox{var}[\hat{F}]
	&= \frac{\mbox{var}[f(\mathbb{Z})]}{N_Z} +  	\sum_{l_1=1}^{N_Z}\sum_{{\substack{l_2=1\\l_1\neq l_2}}}^{N_Z}\frac{\mbox{cov}[f(\mathbb{Z}^{\boldsymbol{\alpha}^{(l_1)}}),f(\mathbb{Z}^{\boldsymbol{\alpha}^{(l_2)}})]	}{N_Z^2} \label{eq:FhatVar}
	\end{align}
	where $\mbox{cov}[x,y] = \langle x y \rangle - \langle x \rangle \langle y \rangle$ denotes the covariance between $x$ and $y$. The first term in Eq.~\eqref{eq:FhatVar} represents the usual convergence in the mean as governed by the law of large numbers \cite{Revesz2014} and hence corresponds to the variance in the sample mean, $\hat{F}$, for on-the-fly calculations. Additional fluctuations in $\hat{F}$, however, can arise when different samples of $\mathbb{Z}^{\boldsymbol{\alpha}}$ are, to some degree, correlated, as embodied in the second term of Eq.~\eqref{eq:FhatVar}. When the size of the pool $\mathcal{P}$ is finite, specific samples of $\mathbb{X}^{(\alpha_p^{(l)})}$ can be present in different realisations of $\mathbb{Z}^{\boldsymbol{\alpha}}$ implying that these correlations are indeed non-zero when the bootstrapping approach is used. 
	
	To assess the strength of these additional fluctuations, we again consider the simple case for which $f(\mathbb{Z}) = \mathbb{Z}$. We therefore seek to assess the covariance matrix, $\mathbb{C}$, for two ordered matrix products $\mathbb{Z}^{\boldsymbol{\alpha}^{(1)}}$ and $\mathbb{Z}^{\boldsymbol{\alpha}^{(2)}}$. Element-wise, we therefore wish to evaluate
\begin{align}
	C_{mn}^{jk}=\langle Z_{jk}^{\boldsymbol{\alpha}^{(1)}}Z_{mn}^{\boldsymbol{\alpha}^{(2)}}\rangle - \langle Z_{jk}^{\boldsymbol{\alpha}^{(1)}}\rangle\langle Z_{mn}^{\boldsymbol{\alpha}^{(2)}}\rangle. \label{eq:C_def}
\end{align}
 For simplicity and without loss of generality, we assume that the first sequence of sampled matrices is $\boldsymbol{\alpha}^{(1)} = (1,2,\ldots,N)$ whereas, the second sampling sequence $\boldsymbol{\alpha}^{(2)}$ is arbitrary, albeit each element is unique since we have assumed $\mathcal{P}$ is sampled without replacement to avoid estimator bias (relaxation of this assumption is discussed in Section~\ref{sec:generalisations}). Accordingly we drop the $\boldsymbol{\alpha}^{(1)}$ dependence of Eq.~\eqref{eq:C_def} and let $ \boldsymbol{\alpha}^{(2)} = \boldsymbol{\alpha}$. 
 Expressing the matrix products of Eq.~\eqref{eq:Z_def} as sums over components, we can explicitly write the element-wise covariance in the form
	\begin{align}
		C_{mn}^{jk} &=  \sum_{\substack{u_1, \ldots, u_{N-1} \\ v_1, \ldots, v_{N-1}}} \left[ \left\langle  \prod_{q=1}^N  X_{u_{q-1}u_q}^{(q)} X_{v_{q-1}v_q}^{(\alpha_q)}  \right\rangle \nonumber \right.\\
   & \quad\quad\quad- \left.\left\langle\prod_{p=1}^N X_{u_{p-1}u_p}^{(p)} \right \rangle \left\langle
		 \prod_{q=1}^N X_{v_{q-1}v_q}^{(\alpha_q)} \right\rangle \right]
		\label{eq:Cjkmn1}
	\end{align}
where again for notational convenience we let  $v_0 = m$ and $v_N=n$ and sums over $v_1,\ldots,v_{N-1}$ run from 1 to $M$. With the sampling assumptions given above and further assuming that each $\boldsymbol{\alpha}$ is equally likely, the probability that the two samples of $\mathcal{P}$ contain at least one common matrix is $P(\{\alpha_j^{(1)}\}\cap\{\alpha_j^{(2)} \} \neq \emptyset) = 1- \binom{N_X-N}{N} / \binom{N_X}{N}$, where $\left(\substack{n\\k}\right)$ denotes the binomial coefficient. Noting, as before, then that samples of $\mathbb{X}$ used to generate $\mathcal{P}$, are taken independently and are by definition identically distributed, we can match any pairs of $X^{(q)}$ and $X^{(\alpha_k)}$ in Eq.~\eqref{eq:Cjkmn1} for which $q=\alpha_k$. A matching pair will generate a factor of the form $\langle X^{(\alpha_q)}_{u_{(\alpha_q -1)} u_{\alpha_q}}  X^{(\alpha_q)}_{v_{q-1}v_q}\rangle$ in the first term of Eq.~\eqref{eq:Cjkmn1} and implies that the matrix $\mathbb{X}^{(\alpha_k)}$ is common to both samples of the matrix pool $\mathcal{P}$. Remaining unpaired matrices however decouple and reduce to a product of their means. To facilitate notation, we can momentarily neglect the ordering of the elements of $\boldsymbol{\alpha}$ and think of it as an integer set. We then introduce the set function $\xi_{\boldsymbol{\alpha}}$ associated with $\boldsymbol{\alpha}$ defined such that $\xi_{\boldsymbol{\alpha}}(\{\alpha_{q_1}\hdots\alpha_{q_k}\}) = \{q_1,\hdots,q_k\}$. In words, given some collection of elements of $\boldsymbol\alpha$, $\xi_{\boldsymbol\alpha}$ picks out the indices corresponding to those elements. Defining now the sets $\mathcal{M} = \xi_{\boldsymbol\alpha}(\{1 ,\hdots,N \} \cap \boldsymbol\alpha)$, $\mathcal{N} = \{ 1,\hdots,N\}\setminus \boldsymbol\alpha$ and $\mathcal{N}' =\xi_{\boldsymbol\alpha}(\boldsymbol\alpha \setminus \{ 1,\hdots,N\}) $, where $\setminus$ denotes the set difference, we find that Eq.~\eqref{eq:Cjkmn1} can be written as
\begin{widetext}
\begin{align}
	C_{mn}^{jk} & =  \sum_{\substack{u_1, \ldots, u_{N-1} \\ v_1, \ldots, v_{N-1}}} \left[
	\prod_{q \in \mathcal{M}} \langle X^{(\alpha_q)}_{u_{(\alpha_q -1)} u_{\alpha_q}}  X^{(\alpha_q)}_{v_{q-1}v_q}\rangle - \prod_{q\in \mathcal{M}}  \langle X_{u_{q-1}u_q}^{(q)} \rangle 
 \langle X_{v_{q-1}v_q}^{(\alpha_q)} \rangle   \right] \prod_{q \in \mathcal{N} } \langle X^{(p)}_{u_{(p-1)} u_{p}}\rangle\prod_{q \in \mathcal{N}' }\langle X^{(\alpha_p)}_{v_{p-1}v_p}\rangle.
	\label{eq:prod_corrs}
\end{align}
\end{widetext}

	 Eq.~\eqref{eq:prod_corrs} is valid quite generally, i.e., for arbitrary $p_{\mathbb{X}}(
  \mathbb{X})$, however, to proceed further we must make same restrictions on the statistical properties of $\mathbb{X}$. As a simple case, we assume that individual transfer matrices, $\mathbb{X}$, are defined by a maximum entropy (i.e., Gaussian) ensemble \cite{Mello1988b}, as is, for example, appropriate for scattering in the diffusive regime. For this model, $\langle \mathbb{X} \rangle = \mathbb{O}$ and
	 $\langle X_{ab}^{(q)} X_{cd}^{(q)} \rangle= r^2 \delta_{ac}\delta_{bd}$ where $\mathbb{O}$ is the null matrix and  $\delta_{jk}$ is the Kronecker delta function, i.e., each individual element of $\mathbb{X}$ has equal variance, whereas distinct elements are uncorrelated.   Whilst this model neglects more complex correlations that may exist in transfer matrices, such as the memory effect, physics imposed symmetry constraints or reflection-transmission correlations \cite{Berkovitsa1994,Osnabrugge2017,Byrnes2021b, Fayard2018}, it serves to sufficiently capture boot-strapping induced correlations that may arise in cascaded models. Moreover, in practice, constituent transfer matrices used in  matrix chained models are frequently drawn from simpler matrix ensembles such as the circular or Gaussian ensembles \cite{Crisanti, Mello1988b,IpsenThesis}. The ballistic scattering regime is discussed further below in Section~\ref{sec:generalisations}. With these assumptions $C^{jk}_{mn}$ is only non-zero when $\boldsymbol{\alpha}$ is an $N$-permutation of $(1,2,\ldots, N)$, that is when the sets of sampled matrices $\mathcal{N}$ and $\mathcal{N}'$ are empty and $\mathcal{M} = \{ 1,\hdots, N \}$, such that we can again enumerate over $q$ whereby
	 \begin{align}
	 	C_{mn}^{jk} & =  \sum_{\substack{u_1, \ldots, u_{N-1} \\ v_1, \ldots, v_{N-1}}}
	 	\prod_{q=1}^N \langle X^{(\alpha_q)}_{u_{(\alpha_q -1)} u_{\alpha_q}}  X^{(\alpha_q)}_{v_{q-1}v_q}\rangle . \label{eq:prod_corrsv2}
	 \end{align}
	 The summations over $v_1,\ldots,v_{N-1}$ appearing in Eq.~\eqref{eq:prod_corrsv2} can be performed analytically yielding
	\begin{align}
		C_{mn}^{jk} &= r^{2N} \!\!\!\sum_{u_1, \hdots, u_{N-1}}\!\!\! \delta_{u_{(\alpha_1 - 1)} v_0} \delta_{v_Nu_{\alpha_N}} 
		\prod_{q=1}^{N-1} \delta_{u_{(\alpha_{(q+1)} - 1 )} u_{\alpha_q}} \label{eq:Cjkmn2}.
	\end{align}
	
	To help evaluate $C_{mn}^{jk}$ we define the multiset $\beta$ of all indices appearing in Eq.~\eqref{eq:Cjkmn2} (allowing for repeated indices) such that $\beta = \{u_{\alpha_1 - 1}, v_0, u_{\alpha_2 - 1} , u_{\alpha_1}, u_{\alpha_3 - 1}, u_{\alpha_2}, \ldots, u_{\alpha_N}, v_N  \}$, which has cardinality $|\beta| = 2(N+1)$, corresponding to the $N+1$ different Kronecker delta functions appearing in Eq.~\eqref{eq:Cjkmn2}. The elements $\{u_0,u_N,v_0,v_N\}$ each have a multiplicity of 1 in $\beta$, whereas $\{u_1,u_2,\ldots,u_{N-1}\}$ have multiplicities of 2. Mirroring the structure of Eq.~\eqref{eq:Cjkmn2} we further define the multisets $\gamma =\{u_{\alpha_1 - 1}, v_0 ,u_{\alpha_N}, v_N   \}$ and $\zeta = \{u_{\alpha_2 - 1} , u_{\alpha_1}, u_{\alpha_3 - 1}, u_{\alpha_2} \ldots\}$ such that $\beta = \gamma \cup \zeta$.  Note that $u_0 \notin \zeta$ if and only if $\alpha_1 -1=0$. Similarly, $u_N \notin \zeta$ if and only if $\alpha_N = N$. 
	With these definitions, consider then the form of Eq.~\eqref{eq:Cjkmn2} when $\{u_0, u_N\} \notin \zeta$ which reduces to (recalling that we previously set $u_0 = j$ etc. for convenience)
	\begin{align}
		C_{mn}^{jk} &= r^{2N} \delta_{jm}\delta_{kn}\sum_{u_1, \hdots, u_{N-1}} \prod_{q=1}^{N-1} \delta_{u_{(\alpha_{(q+1)} - 1 )} u_{\alpha_q}}.
	\end{align}
	The indices appearing in the summations are those contained in  $\{u_1,u_2,\ldots,u_{N-1}\}$ each with multiplicities of two. We can thus group together indices that form disjoint closed cycles. For example, consider the permutation $(1)(2,3,5)(4)$ which we have written in standard cycle notation \cite{Bona2012}. The summation  in this case would take the form
	\begin{align}
		\left[\sum_{u_1} \delta_{u_1u_1} \right]\left[\sum_{u_2}\sum_{u_3}\sum_{u_5} \delta_{u_2u_3}\delta_{u_3u_5}\delta_{u_5u_2} \right] \left[\sum_{u_4} \delta_{u_4u_4} \right]\nonumber
	\end{align}
	where the indices are grouped according to the cycle structure. Given that each index is summed from $1$ to $M$ it is quickly seen that each disjoint set of summations totals $M$ such that
	\begin{align}
		C_{mn}^{jk} =  r^{2N} \delta_{jm}\delta_{kn} M^K \label{eq:Cjkmn_final}
	\end{align}
	where $K$ is the number of disjoint cycles of indices in the sum (which we note do not contain $u_0$ or $u_N$).
	
	Moving now to a more complex case whereby we allow $u_0 \in \zeta$, but still restrict $u_N \notin \zeta$ we have 
	\begin{align}
		C_{mn}^{jk} &= r^{2N} \delta_{kn}\!\!\! \sum_{u_1, \ldots, u_{N-1}}\!\!\!\!\delta_{u_0u_{\alpha_p}} \delta_{v_0 u_{\alpha_1 - 1}}\!\!\prod_{\substack{q=1\\q\neq p}}^{N-1} \delta_{u_{(\alpha_{(q+1)} - 1 )} u_{\alpha_q}},
	\end{align}
 where, since $u_0 \in \zeta$, we note $p$ is the index such that $\alpha_{p+1} = 1$ where $p\neq 0$. We also have that $\alpha_1 -1 >0$. We can once again group indices into summations over closed disjoint cycles of indices, which requires that the cycles do not contain $u_0$ or $v_0$. Each set of indices that can be so factored, contributes a multiple of $M$ to $C_{mn}^{jk}$. We denote the remaining multiset of indices $\psi = \{u_0, u_{\alpha_p},u_{\alpha_1 -1}, v_0 ,u_{\alpha_p }, u_{\alpha_1 -1}, u_f, u_g, \ldots\}$ where $u_f$ and $u_g$ denote general indices which will have multiplicity of two. From the structure of the indices appearing in Eq.~\eqref{eq:Cjkmn2}, it however follows that indices with multiplicity of 2 appear as different arguments to the Kronecker delta functions, i.e., the remaining summation can be written
	\begin{align}
		\sum_{u_{\alpha_p}}\sum_{u_{\alpha_1 -1}}\cdots \sum_{u_f}  \delta_{u_0 u_{\alpha_p}} \delta_{u_{\alpha_p} u_f} \ldots \delta_{u_f u_{\alpha_1-1}} \delta_{u_{\alpha_1-1} v_0}=\delta_{jm}.\nonumber
	\end{align}
	We thus find that $C_{mn}^{jk}$ is again given by Eq.~\eqref{eq:Cjkmn_final}. Analogous arguments show the same to be true for the case that $u_0 \notin \zeta$, but $u_N \in \zeta$. The remaining case ($\{u_0,u_N\} \in \zeta$) can be considered in a similar manner, factoring out disjoint cycles of indices not containing $u_0, u_N, v_0$ and $v_N$, each of which contributes a factor of $M$. The remaining indices reduce upon summation to factors of the form $\delta_{jk}\delta_{mn}$ or $\delta_{jm}\delta_{kn}$ depending on the initial permutation, such that $C_{mn}^{jk}$ is given either by Eq.~\eqref{eq:Cjkmn_final} or by
	\begin{align}
		C_{mn}^{jk} =  r^{2N} \delta_{jk}\delta_{mn} M^K . \label{eq:Cjkmn_final2}
	\end{align}
	
	Given these results we observe that the elements of the correlation matrix formed from $C_{mn}^{jk}$ are either zero or a fixed value ($=r^{2N} M^K$) for any given permutation $\boldsymbol{\alpha}$ (note $\mathbb{C}$ is not diagonal). Moreover, the number of non-zero elements is  
	\begin{align}
		\sum_{j,k,m,n=1}^M \delta_{jk}\delta_{mn} = \sum_{j,k,m,n=1}^M \delta_{jm}\delta_{kn} = M^2 
	\end{align}
	for all possible permutations. Returning to Eq.~\eqref{eq:FhatVar} for $f(\mathbb{Z}) = \mathbb{Z}$, and, considering the total variance in our estimate, as shall be quantified by taking the 1-norm (denoted $\|\ldots\|_1$), we can write
    \begin{align}
    \| \mbox{var}[\hat{\mathbb{Z}}]\|_1 = \frac{\|\mbox{var}[\mathbb{Z}]\|_1}{N_Z}+\sum_{l_1=1}^{N_Z}\sum_{{\substack{l_2=1\\l_1\neq l_2}}}^{N_Z}\frac{C(K,M,N)}{N_Z^2}\label{eq:1normvar}
    \end{align}
 where $K$ is a function of the specific sampled sequences $\boldsymbol{\alpha}^{(l_1)}$ and $\boldsymbol{\alpha}^{(l_2)}$ (or equivalently the sequences $(1,2,\ldots,N)$ and $\boldsymbol{\alpha}^{(l)}$ under a suitable transformation) and 
\begin{align}
    C(K,M,N) = \|\mathbb{C}\|_1 = \!\!\sum_{j,k,m,n=1}^M C_{mn}^{jk} = r^{2N}  M^{K+2}.  
    \label{eq:CNK}
\end{align}
The total covariance $C(K,M,N)$ hence characterises the degree to which $\mathbb{Z}^{(1,\ldots,N)}$ and $\mathbb{Z}^{\boldsymbol{\alpha}}$ are correlated for a given permutation $\alpha$ (note that with a slight abuse of notation we shall use $\alpha$ to denote the permutation described by the sequence $\boldsymbol{\alpha}$). Naturally $C(K,M,N)$ is largest when $\alpha$ is the identity (whereby $K=N$) i.e., $\mathbb{Z}^{\boldsymbol{\alpha}} = \mathbb{Z}^{(1,\ldots,N)}$. Observing further that $\|\mbox{var}[\mathbb{Z}]\|_1 = C(N,M,N)$,  the relative magnitude of each contribution in the second term of Eq.~\eqref{eq:1normvar} with respect to the first term is $\rho / N_Z$, where  
	\begin{align}
		\rho = \frac{C(K,M,N)}{C(N,M,N)} = \frac{1}{M^{N-K}} \label{eq:corr}
	\end{align}
is the Pearson's correlation coefficient, which decreases with matrix size $M$ since $K\leq N$. For cascaded transfer matrix models the correlations between different matrix products hence become less significant for systems with large numbers of scattering modes, therefore promoting use of bootstrapping techniques in such scenarios. Likewise as the number of matrices $N$ within each cascade increases, so the potential correlations between samples decrease in strength. 

From a practical standpoint, it may also be of interest to consider the additional number of  realisations of $\mathbb{Z}$ that must be generated using the bootstrapping method, as compared to the on-the-fly approach, so as to produce the same total estimator variance $\| \mbox{var}[\hat{\mathbb{Z}}]\|_1$. To answer this we recall Eq.~\eqref{eq:1normvar} and note that if only the first term is considered the result corresponds to the total variance of $\hat{\mathbb{Z}}$ for on-the-fly calculations. Therefore, denoting the total number of realisations required to produce a given and fixed $\| \mbox{var}[\hat{\mathbb{Z}}]\|_1$ using a bootstrapped or on-the-fly approach explicitly as $N_Z^{\txtpow{bs}}$ and $N_Z^{\txtpow{otf}}$, it follows (by equating Eq.~\eqref{eq:1normvar} for the two cases and solving the resulting quadratic equation in $N_Z^{\txtpow{bs}}$) that 
\begin{align}
\frac{N_Z^{\txtpow{bs}}}{N_Z^{\txtpow{otf}}} \approx 1 + \sum_{l_1=1}^{N_Z^{\mbox{\tiny{bs}}}}\sum_{{\substack{l_2=1\\l_1\neq l_2}}}^{N_Z^{\mbox{\tiny{bs}}}} \frac{1}{M^{N-K}} ,\label{eq:real_frac}
\end{align}
where we have also used Eq.~\eqref{eq:corr} and assumed that  $\sum_{l_1}\sum_{l_2\neq l_1} \rho \ll N_Z^{\txtpow{otf}}$. It therefore follows that the additional number of realisations required to produce a desired variance also decreases with increasing matrix size $M$ and number of matrices $N$.  

Finally, we consider how the results given generalise when the constituent matrices $\mathbb{X}$ are complex. Notably, all results pertaining to estimator bias presented in Section~\ref{sec:bias} are unchanged. When considering estimator variance, for proper complex random matrices $\mathbb{X}$  with zero mean and equal variance, i.e., $\langle \mathbb{X} \rangle = \mathbb{O}$ and
	 $\langle X_{ab}^{(q)} X_{cd}^{(q)*} \rangle= r^2 \delta_{ac}\delta_{bd}$, formally identical results to those given in Section~\ref{sec:var} can also be derived, albeit using complex generalisations to the correlation functions (such as Eq.~\eqref{eq:C_def}), in which the second matrix factor is conjugated. Note that pseudo-correlation functions are identically zero for proper random variables, i.e., $\langle X_{ab}^{(q)} X_{cd}^{(q)} \rangle= 0$,  and $\mbox{var}[\mbox{Re}(\mathbb{X})] = \mbox{var}[\mbox{Im}(\mathbb{X})] = \mbox{cov}[\mathbb{X},\mathbb{X}^*]/2$ \cite{ScharfComplexBook}. For improper $\mathbb{X}$ the pseudo-correlation would, however, be non-zero requiring further analysis, for which specification of both $\langle X_{ab}^{(q)} X_{cd}^{(q)} \rangle$ and $\langle X_{ab}^{(q)} X_{cd}^{(q)*} \rangle$ would be needed. 

	\section{Enumeration of permuted matrix correlations }\label{sec:enum}

	\newcolumntype{C}{>{\centering\arraybackslash}p{1.5cm}}
	\begin{table*}[t]
		\centering
		\begin{tabular}{|c||C|C|C|C|C|C|C|C|C|C|}		
			\hline
			\backslashbox{N}{K}& 0 & 1 & 2 & 3 & 4 & 5 & 6 & 7 &8 & 9\\
			\hline
			\hline
			1 & 1 		& -			&-		 &-		  &-		&-		&-		&-		&-		&-\\
			2 & 1 		& 1 		&-		 &-		  &-		&-		&-		&-		&-		&-\\
			3 & 3 		& 2 		& 1		 &-		  &-		&-		&-		&-		&-		&-\\
			4 & 8 		& 12 		& 3		 & 1	  &-		&-		&-		&-		&-		&-\\
			5 & 40 		& 44 		& 31 	 & 4	  & 1		&-		&-		&-		&-		&-\\
			6 & 180 	& 324 		& 145 	 & 65	  & 5		& 1		&-		&-		&-		&-\\
			7 & 1,260 	& 1,784		& 1,499	 & 370	  & 120		& 6		& 1		&-		&-		&-\\
			8 & 8,064 	& 16,288 	& 9,772	 & 5,180	  & 805		& 203	& 7 	& 1		&-		&-\\
			9 & 72,576   & 120,672  	& 113,868 & 39,032  & 14,833	& 1,568  & 322	& 8		& 1		&- \\
			10 & 604,800 & 1,327,680 	& 958,956 & 570,044 & 126,861 	& 37,149 & 2,814  & 486 	& 9 	& 1\\
			\hline

		\end{tabular}
		\caption{Table of frequency $\nu({N,K})$ of each value of $C(N,K)$ for all permutations $\alpha_N \in S_N$. \label{tab:numCNKresults}}
	\end{table*}
	
	In addition to quantifying the total correlation between permuted matrix products for a given permutation, it is also relevant to enumerate the frequency with which each value of correlation occurs across the standard group $S_N$ and thus how often each would appear in the summation of Eqs.~\eqref{eq:1normvar} and \eqref{eq:real_frac}. To begin to answer this we seek a more convenient way in which to determine $K$. To do so we revisit the form of Eq.~\eqref{eq:Cjkmn2} and define ordered $(N+1)$ tuples comprising of the indices found in the first and second position of each Kronecker delta respectively, i.e., $\mathbf{p}_1 = (u_{\alpha_1 -1} , u_{\alpha_2 -1} ,\ldots,  u_{\alpha_{N} -1}  , v_N  ) $ and $\mathbf{p}_2 = ( v_0 , u_{\alpha_1} ,  u_{\alpha_2}, \ldots, u_{\alpha_{N-1}} , u_{\alpha_{N}} )$. Noting that $\alpha_q -1$ spans $[0,N-1]$  and $\alpha_q$ spans $[1,N]$ for $q\in[1,N]$, it follows that the subscripts are unique within each tuple and we can equivalently consider the subscripts themselves whereby $\mathbf{p}_1 = ({\alpha_1 -1} , {\alpha_2 -1} ,\ldots  {\alpha_{N} -1}  , N  ) $ and $\mathbf{p}_2 = (0 , {\alpha_1} ,  {\alpha_2}, \ldots {\alpha_{N-1}} , {\alpha_{N}} ) $. The cycles of the $(N+1)$-permutation $\sigma$ that maps $\mathbf{p}_2 \rightarrow \mathbf{p}_1$ (i.e., considering $\sigma$ as a permutation matrix $\mathbf{p}_1^T = \sigma \mathbf{p}_2^T$) therefore correspond to the cycles of indices in which we are interested. Specifically, $K$ is the number of disjoint cycles of $\sigma$ which do not contain $0$ or $N$, whereby it follows that $K\leq N-1$.

	To express $\sigma$ in terms of $\alpha$ we let $\mathbf{p}_0 = (0,1,2,\ldots,N)$ and define the $(N+1)\times(N+1)$ permutation matrices $\sigma_1$ and $\sigma_2$ whereby $\mathbf{p}_k^T = \sigma_k\mathbf{p}_0^T$ for $k = 1,2$.
	We thus immediately see that $\sigma_2$ has the structure
	\begin{align}
		\sigma_2 = \left[\begin{array}{cc} 
			1& \mathbf{0}^T \\
			\mathbf{0} & \sigma_{\alpha}  
		\end{array}\right]\label{eq:sigma2_structure}
	\end{align}
	where $\sigma_\alpha$ is the $N\times N$ permutation matrix associated with $\alpha$ and $\mathbf{0}$ is an $N$ element column vector of zeros.  We also note  $\mathbf{p}_1^T = \sigma_1 \sigma_2^{-1}\mathbf{p}_2^T$ such that $\sigma = \sigma_1 \sigma_2^{-1}$. Defining the cyclic-shifting operator
	\begin{align}
		\sigma_+ = \left[\begin{array}{cccccc} 
			0& 0& \ldots& 0 & 0 & 1\\
			1& 0& \ldots& 0 & 0 & 0\\
			0& 1& \ldots& 0 & 0 & 0 \\
			\vdots & \vdots & \ddots & \vdots &\vdots & \vdots\\
			0 & 0 & \ldots&1 & 0 & 0 \\
			0 & 0 & \ldots& 0 & 1 & 0 
		\end{array}\right],
	\end{align}
	which shifts elements in a tuple $p$ to the right in a cyclic manner, and noting that shifting to the left (as performed by $\sigma_- = \sigma_+^{-1} = \sigma_+^T$) is equivalent to subtraction of unity modulo $N+1$, we can write 
	\begin{align}
		\sigma_1 = \sigma_- \sigma_2 \sigma_+ = \left[\begin{array}{cc} 
			\sigma_{\alpha}&\mathbf{0} \\
			\mathbf{0}^T &1
		\end{array}\right] ,\label{eq:sigma1_structure}
	\end{align}
	yielding
	\begin{align}
		\sigma =  \sigma_- \sigma_2 \sigma_-^{-1} \sigma_2^{-1}. \label{eq:sigma_product_form}
	\end{align}
	From Eq.~\eqref{eq:sigma_product_form} it follows simply that $\mbox{det}[\sigma] = 1$ implying that $\sigma$ is always an even permutation \cite{Bona2012}. Furthermore, since $\sigma$ cannot map the 1st element of $\mathbf{p}_2$ (which is $0$ by construction) to the last element of $\mathbf{p}_1$ (which is $N$), it follows that the $(N+1,1)$th element of $\sigma$ is zero, i.e., $\sigma_{N+1,1} = 0$.

	With this formalism we can thus more simply numerically enumerate the frequency of $C(K,M,N)$ across all possible permutations $\alpha_N \in S_N$, which we denote $\nu({K,N})$. Note, the argument $M$ has been omitted since the frequency derives purely from combinatorial aspects of matrix products and is hence independent of matrix size. Results are presented in Table~\ref{tab:numCNKresults} for small values of $N$. We note the following results (which have been numerically verified, via exhaustion, up to $N=13$):
	\begin{align}
	   \nu({N-1,N}) &= 1 ,  \label{eq:seq1} \\
	    \nu({N-2,N}) &= N-1, \label{eq:seq2} \\
		\nu({N-3,N}) &= \binom{N-1}{N-3} + \binom{N}{N-3} + \binom{N+1}{N-3}, \label{eq:seq3} \\
		\nu(0,N) &= N! / \lfloor (N+2)/2 \rfloor, \label{eq:seq4}
	\end{align}
	for all $N$. We observe Eqs.~\eqref{eq:seq3} and \eqref{eq:seq4} correspond to integer sequences A005718 and A107991 respectively \cite{oeisA005718, oeisA107991}. Mathematical proofs of Eqs.~\eqref{eq:seq1}--\eqref{eq:seq4} are presented in Appendix~\ref{appendix} in turn. Eqs.~\eqref{eq:seq1}-\eqref{eq:seq3} we prove through exhaustion and conjecture that the graphical approach employed in proof of Eq.~\eqref{eq:seq3} can be extended to other cases. For Eq.~\eqref{eq:seq4} we however present an alternative approach based on establishing a bijection to known combinatorial results. 
	
	The results of Table~\ref{tab:numCNKresults} and the analytic formulae for $\nu(K,N)$ highlight a number of important trends with respect to correlations between cascaded transfer matrices. Most importantly, it is evident that the relative frequency
    \begin{align}
    \nu_{rel}=\nu(K,N)/N! \label{eq:relfreq}
    \end{align} 
    of the higher correlation cases (i.e., smaller $S=N-K$) across the standard group becomes smaller for larger $N$. Consequently, the majority of thick media transfer matrices found from cascading large numbers of thin-section transfer matrices are only weakly correlated with each other, if at all. Correlations thus decrease in both relative magnitude and relative frequency as the number of matrices in a given product increases and consequently the additional fluctuations introduced in an estimator $\hat{F}$ will also typically be lower. The frequency function $\nu(K,N)$ is nevertheless peaked at a non-zero correlation, showing a non-monotonic dependence on $K$ for a fixed $N$.

\section{Routes to generalisation}\label{sec:generalisations}
	
In the above analysis we made a number of restrictive assumptions. We now briefly discuss and outline how generalisation of our results could be sought by the interested reader.    
 
 Firstly, we note that the statistical model assumed thus far for the thin section transfer matrix $\mathbb{X}$, was appropriate for the diffusive scattering regime. Alternatively, we can consider the case for which a thin slab only weakly perturbs the field incident upon it, as is more relevant for calculations in the ballistic regime and DMPK type models. For weakly scattering slabs the transfer matrix can be written in the form $\mathbb{X} = \mathbb{I} + \Delta \mathbb{X}$ where $\Delta \mathbb{X}$ describes the scattering based perturbation from the identity matrix $\mathbb{I}$. As a simple model we can assume now $\langle\Delta \mathbb{X}\rangle = \mathbb{O}$ where $\mathbb{O}$ is the null matrix and $\langle \Delta\mathbb{X}_{ab} \Delta\mathbb{X}_{cd} \rangle = r^2 \delta_{ac}\delta_{bd}$ where $\delta_{jk}$ is the Kronecker delta function. It immediately follows that 
		 $\langle {X}_{ab} {X}_{cd} \rangle = \delta_{ab}\delta_{cd} + r^2 \delta_{ac}\delta_{bd}$ such that each individual element of $\mathbb{X}$ has equal variance, whereas distinct off-diagonal elements are uncorrelated. Note, diagonal elements of the transfer matrix $\mathbb{X}$ describing direct transmission possess a non-zero mean in contrast to the diffusive result. 
	
  For the case that $\boldsymbol{\alpha}$ is an $N$-permutation of $(1,2,\ldots,N)$ we note that
  \begin{widetext}
  \begin{align}
  	C_{mn}^{jk} & =  \sum_{\substack{u_1, \ldots, u_{N-1} \\ v_1, \ldots, v_{N-1}}}\left[
  	\prod_{q=1}^N \langle X^{(\alpha_q)}_{u_{(\alpha_q -1)} u_{\alpha_q}}  X^{(\alpha_q)}_{v_{q-1}v_q}\rangle- \prod_{q=1}^N  \langle X_{u_{q-1}u_q}^{(q)} \rangle 
  	\langle X_{v_{q-1}v_q}^{(\alpha_q)} \rangle  \right]\label{eq:Cmeso1} \\
  	&= \sum_{\substack{u_1, \ldots, u_{N-1} \\ v_1, \ldots, v_{N-1}}}
\left[  	\prod_{q=1}^N ( \delta_{u_{\alpha_q -1} u_{\alpha_q}} \delta_{v_{q-1}v_q }  + r^2 \delta_{u_{\alpha_q-1} v_{q-1} }\delta_{u_{\alpha_q}v_q} ) - \prod_{q=1}^N   \delta_{u_{q-1}u_q}  \delta_{v_{q-1}v_q}\right]. \label{eq:Cmeso2}
  \end{align}
  \end{widetext}
	Expanding the first product term gives a power series in $r^2$, for which a general term can be written in the form
	\begin{align}
r^{2|\gamma|} \sum_{\substack{u_1, \ldots, u_{N-1} \\ v_1, \ldots, v_{N-1}}} \prod_{q\in \beta}\delta_{u_{\alpha_q -1} u_{\alpha_q}} \delta_{v_{q-1}v_q }  \prod_{p\in \gamma} \delta_{u_{\alpha_p-1} v_{p-1} }\delta_{u_{\alpha_p}v_p},
	\end{align}
where $\beta$ and $\gamma$ are sets of indices dependent on which term in the series we consider and $\beta\cup\gamma = \{1,2,\ldots ,N\}$. Following a similar logic to above, we can sum over $v_1,\ldots,v_{N-1}$  such that a general term in the expansion is given by
\begin{align}
r^{2|\gamma|} \!\!\!\!\sum_{u_1, \ldots, u_{N-1}}  \!\!\! \delta_{u_{\tau_1 -1}v_0} \delta_{u_{\tau_{|\gamma|}}v_N}    \prod_{j=1}^{|\beta|} \delta_{u_{\rho_j -1} u_{\rho_j} } \! \prod_{k=1}^{|\gamma|-1} \delta_{u_{\tau_k} u_{\tau_{k+1}-1}} \label{eq:rexpterm}
\end{align}
where $\rho_j = \alpha_{\beta_j}$ and $\tau_k = \alpha_{\gamma_k}$. Eq.~\eqref{eq:rexpterm} can then in turn be evaluated by counting the cycles of the permutation which transforms the tuple $\mathbf{p}_1$ to $\mathbf{p}_2$ for
 \begin{align}
 \mathbf{p}_1 &= (\rho_1-1,\ldots,\rho_{|\gamma|}-1,\tau_1 - 1,\tau_2-1,\ldots,\tau_{|\tau|}-1,N )\nonumber\\
 \mathbf{p}_2 &=(\rho_1,\ldots,\rho_{|\gamma|},0,\tau_1 ,\ldots,\tau_{|\tau|-1},N ) .\nonumber
 \end{align}
The analysis for a single term in Eq.~\eqref{eq:Cmeso2} can thus proceed in an analogous manner to that given above for the diffusive regime. Given the indices involved in each term, as defined by the sets $\beta$ and $\gamma$ (cf. Eq.~\eqref{eq:rexpterm}), differ for each term, it follows that the relevant permutation and the resulting contribution to the total covariance also varies term by term. Nevertheless, the 
route to the ballistic regime is apparent, if not somewhat tedious. Ultimately, similar trends in the bootstrapped induced fluctuations in estimates of population averages, in terms of matrix size $M$ and number of factors in the matrix products $N$ would result. 

Finally, we briefly consider the generalisation whereby matrices $\mathbb{X}^{(j)}$ are sampled from the pool $\mathcal{P}$ with replacement. As per our earlier discussion in Section~\ref{sec:biasvar}, this sampling strategy can introduce undesirable bias in an estimate of the sample mean, however, is simple to implement. The consequence of sampling from $\mathcal{P}$ with replacement, is that indices in both $\boldsymbol{\alpha}^{(1)}$ and $\boldsymbol{\alpha}^{(2)}$ may be repeated (note it is now necessary to reintroduce the more general notation since the previous assumption that $\boldsymbol{\alpha}^{(1)} = (1,2,\ldots,N)$ does not adequately encompass all possibilities). In calculation of covariance matrix elements (Eq.~\eqref{eq:Cjkmn1}), repeated indices in $\boldsymbol{\alpha}^{(1)}$ and $\boldsymbol{\alpha}^{(2)}$ imply that upon grouping averages of like terms (cf. Eq.~\eqref{eq:Cjkmn2}), one can obtain higher order moments. For instance, if $\boldsymbol{\alpha}^{(1)} = (1,2,2)$ and $\boldsymbol{\alpha}^{(2)}= (4,3,2)$, the first product appearing in Eq.~\eqref{eq:Cjkmn1} would include a factor of the form $\langle X^{(2)}_{pq}X^{(2)}_{rs}X^{(2)}_{uv}\rangle$ (where we let the subscripts take arbitrary values for simplicity). Use of the moments-cumulant formula \cite{Leonov1959} (or for Gaussian random variables, Isserlis' theorem) would in principle allow such higher order moments to be expressed in terms of lower orders. So doing, however, requires careful attention to be paid to the partitioning of the index sets for each individual case, which rapidly becomes cumbersome, but is in principle possible. 

\section{Conclusions}
In this article we have considered the problem of simulation of wave propagation in random media using random matrices. Through forming the appropriate ordered product of random matrices drawn from a suitable ensemble, waves can in principle be propagated through thick media. Indeed, such cascaded models have spawned a number of interesting insights into the statistics of, for instance, the eigenvalues and mode spacings of the underlying ensembles \cite{Eynard1998, Mahoux1998,Akeman2013,Mehta1994,Furstenberg1960, Crisanti, Akeman2023}. To improve efficiency of cascaded matrix models, we have here proposed a bootstrapped dual pool approach whereby after pre-generation of pools of random transfer matrices, many realisations can be simulated through matrix resampling. The proposed approach not only reduces the number of matrix generations required, but also reduces the total number of matrix products, which can limit computation times when there are many scattering modes available. We have however also shown, that matrix resampling inherent in the proposed technique,  can in principle adversely affect the statistical properties of the resulting ensemble. In particular, we considered how bootstrapping can introduce undesirable statistical bias and additional variance in estimates of  population averages of properties such as transmission. Whilst bias was shown to be avoidable through an appropriate sampling strategy, correlations that arise from matrix reuse, and the additional estimator covariance that follows, remain. The strength of this additional covariance, as characterised by Eq.~\eqref{eq:1normvar}, was found to be dependent on the correlation between different matrix permutations. An extensive study of the magnitude and frequency of such correlations was thus also presented, including proof of a number of closed form analytic formulae (Eqs.~\eqref{eq:seq1}--\eqref{eq:seq4}). An important finding of this study was that the consequences of spurious correlations will typically decrease in severity as the dimension of random matrices and the number of matrix products increases (see Eqs.~\eqref{eq:corr} and \eqref{eq:relfreq}). Fortunately, bootstrapping approaches are only required in the regime where scattering channels are numerous and scattering media are many mean free paths in length. Our results therefore show that in spite of the detrimental statistical effects of bootstrapping, practical application of either a single or dual pool approach is not limited by them.

\acknowledgements
NB was supported by Nanyang Technological University grant 
 number SUG:022824-00001. GRWG was supported by the Singapore Ministry of Education Academic Research Fund (Tier 1); grant numbers: RG21/20 and RG23/20. MRF was supported by funding from the Institute for Digital Molecular Analytics and Science (IDMxS) under the Singapore Ministry of Education Research Centres of Excellence scheme.

 \appendix

 	\section{Mathematical  proofs of correlation results}\label{appendix}

	\subsection{Proof of Eq.~\eqref{eq:seq1}} 
	To prove Eq.~\eqref{eq:seq1} we begin by considering the cycle type of $\sigma$ \cite{Bona2012}. Specifically we recall that an $N+1$ permutation with $a_i$ cycles of length $i$ has cycle type $\mathbf{a} = (a_1,a_2,a_3,\ldots,a_N,a_{N+1})$. It follows then that $N+1 = \sum_{i} i a_i$ and the total number of cycles $N_c = \sum_i a_i$. Since we are ultimately interested in the number of cycles $K$ that do not contain $0$ or $N$, we must consider two cases: when $0$ and $N$ are in the same cycle whereby $N_c = K+1$ (Case A) and when $0$ and $N$ are in different cycles, whereby $N_c = K+2$ (Case B). Now assuming $K = N-S$ we can write 
	\begin{align}
		K = \sum_i a_i - \Delta = N - S = \left(\sum_i i a_i\right) - 1 - S, \label{eq:proof1}
	\end{align}
	where $\Delta = 1$ for Case~A and $\Delta = 2$ for Case~B. Upon rearrangement, Eq.~\eqref{eq:proof1} becomes 
	\begin{align}
		\sum_i (i-1)a_i  = a_2 + 2 a_3 + \ldots + N a_{N+1}= S - \Delta+ 1. \label{eq:cycleconstraint}
	\end{align}
	For $S = 1$ we can see by inspection that the possible cycle types satisfying Eq.~\eqref{eq:cycleconstraint} for 
	Case~A are $\mathbf{a}_A =  (N-1 , 1 , 0 , 0 , \ldots )$ such that the cycles must be $(0,N)(1)(2)\ldots(N-1)$, whereas for Case~B $	\mathbf{a}_B =  (N+1 , 0 , 0 , 0 , \ldots )$ such that the cycles must be $(0)(1)(2)\ldots(N-1)(N)$. Noting that odd permutations have an even number of even cycles \cite{Bona2012}, the former solution, $\mathbf{a}_A$, is not permitted since it describes an odd permutation. There is thus a single permutation $\sigma = (0)(1)(2)\ldots(N-1)(N)$, corresponding to $\alpha = (1)(2)\ldots(N)$, i.e., the identity, for which $S=1$ therefore completing the proof of Eq.~\eqref{eq:seq1}. 

\begin{table*}[t]
	\centering
	\begin{tabular}{| c||c|c|c|}		
		\hline
		~~~Class~~~ &  ~~Cycle Structure~~  & ~~Constraints~~ & Total Cases\\
		\hline	\hline
		B1.1&	$(0,p)(q,N)$ & $q<p$ & $\frac{1}{2}(N-1)(N-2)$ \\
		B1.2&	$(0,p)(q,r)$ & $p<r<q$ &$\frac{1}{6}(N-1)(N-2)(N-3)$\\
		B1.3&   $(p,q)(r,N)$ & $p<r<q$ & $\frac{1}{6}(N-1)(N-2)(N-3)$\\
		B1.4&   $(p,q)(r,s)$ & $p<s<q<r$  &$\frac{1}{24}(N-1)(N-2) (N-3)(N-4)$\\
		B2.1&   $(0,p,q)$  & $q<p$ & $\frac{1}{2}(N-1)(N-2)$\\
		B2.2&   $(p,q,N)$ & $q<p$ & $\frac{1}{2}(N-1)(N-2)$\\
		B2.3&   $(p,q,r)$ & $p<r<q$ &$\frac{1}{6}(N-1)(N-2)(N-3)$\\
		\hline
	\end{tabular}
	\caption{Index constraints and corresponding number of allowable permutations for different classes of cycle structure. \label{tab:permproofcases}}
\end{table*}

\subsection{Proof of Eq.~\eqref{eq:seq2}} 
	We can approach the proof of Eq.~\eqref{eq:seq2} in a similar manner to that used to prove Eq.~\eqref{eq:seq1} above. Consider then possible solutions of Eq.~\eqref{eq:cycleconstraint} when $S=2$, i.e., $K=N-2$. For Case~A we require
	\begin{align}
		a_2 + 2a_3 + 3 a_4 + \ldots = 2
	\end{align}
	such that 
	\begin{align}
		\mathbf{a}_A =  
		\begin{dcases*}
		 (N-3, 2,0,0,\ldots)  \\  (N-2, 0,1,0,\ldots) 
		\end{dcases*} ,
	\end{align}
	corresponding to cycle structures of the form 
	\begin{eqnarray}
		(0,N)(1)(2)\ldots(p,q)\ldots(N-1)\label{eq:proof2}
	\end{eqnarray}
	and 
	\begin{eqnarray}
		(0,p,N)(1)(2)\ldots(p-1)(p+1)\ldots(N-1) \label{eq:cycletype2}
	\end{eqnarray}
	respectively, where $p\neq q$ and ${p,q}\in [1,N-1]$. Cycle structures of the form of Eq.~\eqref{eq:proof2} are not permitted since $\sigma$ cannot map $0$ directly to $N$ ($\sigma_{N+1,1} =0$), whereas cycle structures of the form of Eq.~\eqref{eq:cycletype2} are permitted, such that there are $N-1$ corresponding permutations. Inspection of the structure of the associated permutation matrices also show that they are of the form required by Eqs.~\eqref{eq:sigma2_structure}--\eqref{eq:sigma_product_form} (this is discussed further in the next section). For Case~B, i.e., $\Delta = 2$, we require
	\begin{align}
		a_2 + 2a_3 + 3 a_4 + \ldots = 1
	\end{align}
	which has the solution $(a_2, a_3,\ldots) = (1,0,\ldots)$ such that $\mathbf{a}_B=(N-1,1,0,0,\ldots)$. This corresponds to an odd permutation implying these solutions are forbidden. In total there are hence $N-1$ allowed permutations of the form of Eq.~\eqref{eq:cycletype2} for which $K = N-2$, hence concluding our proof of Eq.~\eqref{eq:seq2}.	
	
\subsection{Proof of Eq.~\eqref{eq:seq3}} 
	To prove Eq.~\eqref{eq:seq3} we once more consider the possible solutions to Eq.~\eqref{eq:cycleconstraint} now for $S=3$, i.e., $K=N-3$, which are 
	\begin{align}
		\mathbf{a}_{A} = 
		\begin{dcases}
			(N-5,3,0,0,0,\ldots) \\
			(N-4,1,1,0,0,\ldots)\\
			(N-3,0,0,1,0,\ldots)			
		\end{dcases},
	\end{align}
corresponding to odd (and hence invalid) permutations, and 
\begin{align}
	\mathbf{a}_B = \begin{dcases}
		(N-3,2,0,0,0,\ldots)\\
		(N-2,0,1,0,0,\ldots)
		\end{dcases},
\end{align}
which correspond to even, and hence potentially acceptable solutions. Seven distinct classes of cycle structures for the Case B cycle types can be constructed, as are listed in Table~\ref{tab:permproofcases} (omitting one cycles for clarity), where $p,q,r,s \in [1,N-1]$ are all distinct. A naive consideration of the combinatorics of possible values of $p,q,r,s$, however, does not yield Eq.~\eqref{eq:seq3}, as some combinations are inconsistent with the required structure of $\sigma$ as dictated by Eq.~\eqref{eq:sigma_product_form}. To illustrate we use the two-line representation of permutations and consider the ordered action of $\sigma_2^{-1}$ and $\sigma_1$ on the original sequence $[0,1,2,\ldots,N-1,N]$, which can hence be represented using a three-line representation.

\begin{figure*}[t]
	\centering
	\includegraphics[width=0.78\textwidth]{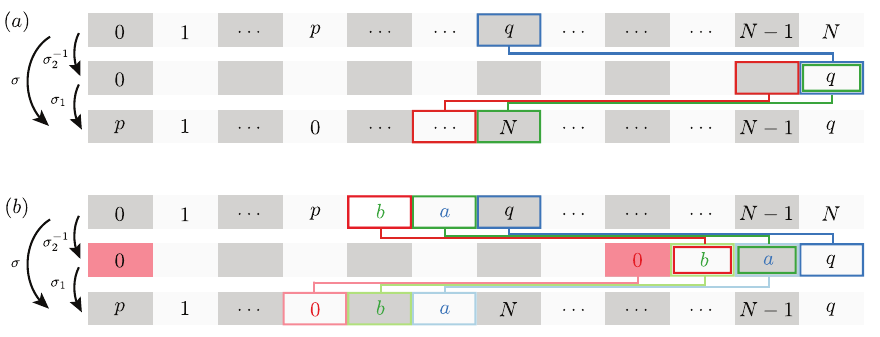}
	\caption{(a) Three line representation of the sequential permutations $\sigma_2^{-1}$ and $\sigma_1$. The existence of the transposition shown in blue in $\sigma_2^{-1}$, which has the inverse shown in green, implies that the shifted inverse transposition shown in red exists in $\sigma_1$. (b) Example graphical proof for Class B1.1 cycle structures. Repeated application of the permutation structure shown in (a) leads to a logical contradiction for $p<q$ as highlighted.}  
	\label{fig:permproof1}
\end{figure*}

Consider the case when the cycle structure of $\sigma$ is of Class B1.1 as an initial illustration. Given that $\sigma_2^{-1}$ does not permute the `0' element (as easily seen from Eq.~\eqref{eq:sigma2_structure}) we can immediately populate the `0' element into the first position on the second row as shown in Figure~\ref{fig:permproof1}(a). Further noting that the cycle structure given in Table~\ref{tab:permproofcases} means that under the action of both permutations $0\rightarrow p$, we can place $p$ into the first element of the third row, whilst simultaneously populating the $p$th element with 0. Similarly, $\sigma_1$ does not affect the last element of the sequence upon which it acts (cf. Eq.~\eqref{eq:sigma1_structure}), and the cycle structure implies $N\rightarrow q$ (and vice versa), whence we can fill additional positions in our three-line representation as also shown in Figure~\ref{fig:permproof1}(a). In Figure~\ref{fig:permproof1}(a), note that we have implicitly assumed that $p<q$. For cycle structures of Class B1.1, all other cycles are one-cycles such that the elements in the upper and lower lines must match for all other columns in our representation (albeit, importantly they need not match in the middle row). To allow for sequences of general length we denote these matching elements using ellipses ($\cdots$). 

Blank spaces in Figure~\ref{fig:permproof1}(a) correspond to elements that do not follow directly from the cycle structure of $\sigma$ and are thus as yet unknown in general, however, we can attempt to deduce them by invoking the transposition structure imposed by Eqs.~\eqref{eq:sigma2_structure} and \eqref{eq:sigma1_structure}. To visualise this structure, assume that $\sigma_2^{-1}$ induces the transposition indicated in blue in Figure~\ref{fig:permproof1}(a), which is inherited from, and encoded in, $\sigma_\alpha^{-1} = \sigma_\alpha^T$ (or equivalently $\sigma_\alpha$). Crucially, $\sigma_1$ also derives from $\sigma_\alpha$ and so it is helpful to consider the inverse of the blue transposition, which we have depicted in Figure~\ref{fig:permproof1}(a) in green. Eq.~\eqref{eq:sigma1_structure} however tells us that to fully describe $\sigma_1$ we must also consider the cyclic operator, which ultimately yields the shifted transposition depicted in red. Individual transpositions arising from the action of $\sigma_1$ can thus be graphically inferred from each transposition in $\sigma_2^{-1}$. 

With these observations we can now attempt to  populate all elements of the three-line representation and exclude any cases from our enumeration which lead to logical contradictions. With reference to Figure~\ref{fig:permproof1}(b) and again considering Class B1.1 for $p<q$, we start by considering the transposition necessary to place $q$ in the last element of the second row (depicted in dark blue). As discussed we can immediately infer that the $\sigma_1$ must contain the transposition shown in light blue (we henceforth use the colour coding whereby a dark colour transposition infers the corresponding light coloured transposition). Noting the existence of the light blue transposition, we can back propagate the element in the final row (here denoted $a$) to the correct position in the second row. Since $a$ belongs to a one-cycle, we can then conclude that $\sigma_2^{-1}$ must contain the transposition shown in dark green, which in turn implies the permutation shown in light green. This pattern repeats (as shown in both Figure~\ref{fig:permproof1}(b) and Supplementary Animation 1 \cite{supp}) until we reach the light red transposition. This transposition requires special attention since it contains the `0' element. In particular, if we back propagate `0' to the middle row, we find that it necessarily does not originate from the first position. This is a contradiction with the structure of $\sigma_2^{-1}$, which dictates that `0' must appear in the first position of the middle row. Therefore, we can conclude that any cycle structure of Class B1.1 with $p<q$ does not admit a permutation of the correct structure. Repeating this process, however starting from the assumption that $p > q$ generates no such contradiction, as shown in Supplementary Animation 1 \cite{supp}, such that these cases are admissible. Counting the number of different possible choices of $p$ and $q$ satisfying this constraint ($p>q$) yields $(N-1)(N-2)/2$ acceptable permutations. An exhaustive application of these rules to the cycle types defined in Table~\ref{tab:permproofcases} and for different possible orderings of $p,q,r,s$ (as applicable) is depicted in the Supplementary Animations \cite{supp}. Ultimately we find the index constraints listed in Table~\ref{tab:permproofcases} along with the corresponding number of possible permutations. In calculation of the number of possible permutations it should be remembered that cycles are unique only up to cyclic permutations. The graphical proof described above, for example, shows that $r < p < q$ and $q < p < r$ admit allowable Class B1.2 permutations, however, these different orderings yield identical permutations since the cycles $(q,r)$ and $(r,q)$ are equivalent. Care must hence be taken not to count such duplicate permutations multiple times. Finally, summation of all possibilities listed in Table~\ref{tab:permproofcases} and some simple algebraic manipulation yields our desired result, in the form of Eq.~\eqref{eq:seq3}. We conjecture such a graphical proof by exhaustion could be extended to other values of $S$, however, we leave this to the enthusiastic reader.

\subsection{Proof of Eq.~\eqref{eq:seq4}} 

Finally, we turn our attention to $S=N$, i.e., $K=0$.
We define $\mathbf e_i$ to be the standard basis vector of the $(N+1)$ dimensional real coordinate space whose $i$th entry is $1$ and the rest of its entries are $0$.
The possible solutions of Eq.~\eqref{eq:cycleconstraint}
are $\mathbf a_A = \mathbf e_{N+1}$ and, for each $i \in \{1,\dots,\lfloor (N+1)/2 \rfloor\}$, the vector $\mathbf a_B = \mathbf e_{i} + \mathbf e_{N+1-i}$.

In our proof of Eq.~\eqref{eq:seq4} we treat odd and even $N$ separately. Consider first then the case where $N=2n$ is even.
The solutions for Case B and even $N$ are immediately seen to be invalid since $\sigma$ is an even permutation.
Consequently, $\sigma$ must have cycle type $\mathbf e_{N+1}$, i.e., it is an $N+1$ cycle.
Observe from Eq.~\eqref{eq:sigma_product_form} that $\sigma$ is expressed as the product of $\sigma_{-}$, which by definition is an $N+1$ cycle, and another permutation of cycle type $\mathbf e_{N+1}$, since $\sigma_2 \sigma_{-}^{-1}\sigma_2^{-1}$ is conjugate to $\sigma_{-}$.
The number of permutations $\pi$ of cycle type $\mathbf e_{N+1}$ such that $\sigma_{-}\pi$ also has cycle type $\mathbf e_{N+1}$ is known to be equal to $(2n)!/(n+1)$~\cite{Stanley,Zagier}.
Hence it immediately follows that
\begin{equation}
    \nu(0,2n) = (2n)!/(n+1),
    \label{eq:even0}
\end{equation}
corresponding to Eq.~\eqref{eq:seq4} for even $N$. 

We now consider the case for which $N= 2n+1$ is odd, for which we find that Case A solutions are invalid due to permutation symmetry.
Following the above logic, we thus wish to count the number of permutations $\pi$ of cycle type $\mathbf e_{N+1}$ such that $\sigma_{-}\pi$ is the product of two disjoint cycles $c_0$ and $c_N$ with $0 \in c_0$ and $N \in c_N$.
We shall let $\Psi_k(N)$ denote the set of permutations $\pi = \sigma_2\sigma_-^{-1}\sigma_2^{-1}$ such that $\sigma_-\pi$ is the product of two disjoint cycles $c_0$ and $c_N$ with $0 \in c_0$, $N \in c_N$, and the cycle $c_0$ consists of $k$ elements.
Furthermore, we let $\Gamma(N)$ denote the set of permutations $\pi$ of cycle type $\mathbf e_{N+1}$ such that $\sigma_{-}\pi$ also has cycle type $\mathbf e_{N+1}$.
First, we observe that there is a bijection from $\Psi_1(N)$ to $\Gamma(N-1)$. 
Indeed, it is straightforward to check that the cycle $(0,\alpha_0,\dots,\alpha_{N-1}) \in \Psi_1(N)$ if and only if $(\alpha_0-1,\dots,\alpha_{N-1}-1) \in \Gamma(N-1)$.
Using Eq.~\eqref{eq:even0}, we therefore find that the cardinality of our sets satisfy
\begin{equation}
    |\Psi_1(N)| = |\Gamma(N-1)| = (2n)!/(n+1).
    \label{eq:even02}
\end{equation}
Next we establish a bijection between the sets $\Psi_k(N)$ and $\Psi_{k+1}(N)$ for $k \in \{1,\dots,N-1\}$.
Supposing that the permutation $\pi = (0,\alpha_0,\dots,\alpha_{N-1}) \in \Psi_k(N)$, we can write
\begin{align}	
\sigma_-\pi = (0,\beta_1,\dots,\beta_{k-1})(\gamma_1,\dots,\gamma_{N-k},N).
\end{align}	
Letting then $g \in \{0,\dots,N-1\}$ be the index such that $\alpha_g = \gamma_{N-k}+1$ and setting
$\pi^{\prime}= (0,\alpha_{g},\alpha_{g+1},\dots, \alpha_{g+N}),$
with indices reduced modulo $N$, it follows that 
\begin{align}	
 \sigma_-\pi^\prime = (0,\gamma_{N-k},\beta_1,\dots,\beta_{k-1})(\gamma_1,\dots,\gamma_{N-k-1},N)
 \end{align}
and hence $\pi^\prime \in \Psi_{k+1}(N)$. This map from $\Psi_k(N)$ to $\Psi_{k+1}(N)$ has an inverse in the following sense.
For $k \ge 1$, if we suppose that the permutation $\pi = (0,\alpha_0,\dots,\alpha_{N-1}) \in \Psi_{k+1}(N)$, we have
\begin{align}	
\sigma_-\pi = (0,\beta_1,\dots,\beta_{k-1},\beta_k)(\gamma_1,\dots,\gamma_{N-k-1},N).
\end{align}	
Letting $g \in \{0,\dots,N-1\}$ be the index such that $\alpha_g = \beta_2+1$ if $k \geq 2$ and $\alpha_g = 1$ if $k = 1$, and furthering setting
$\pi^{\prime}= (0,\alpha_{g},\alpha_{g+1},\dots, \alpha_{g+N-1}),$
with indices reduced modulo $N$, it then follows that 
\begin{align}	
\sigma_-\pi^\prime = (0,\beta_2,\dots,\beta_{k})(\gamma_1,\dots,\gamma_{N-k-1},\beta_1,N),
 \end{align}	
whereby $\pi^\prime \in \Psi_{k}(N)$. 

We have thus established that $|\Psi_{k}(N)| = |\Psi_{k+1}(N)|$ for all 
$k \in \{1,\dots,N-1\}$.
Next noting that we can write $\nu(0,N) = \sum_{k=1}^N |\Psi_{k}(N)|$, we can deduce upon using Eq.~\eqref{eq:even02} that
\begin{equation}
    \nu(0,2n+1) = (2n+1)\frac{(2n)!}{n+1} = \frac{(2n+1)!}{n+1}.
    \label{eq:odd}
\end{equation}
Finally, upon combining Eq.~\eqref{eq:even0} and Eq.~\eqref{eq:odd} we obtain Eq.~\eqref{eq:seq4}.


\begin{thebibliography}{10}
\newcommand{\enquote}[1]{``#1''}

\bibitem{Pamel2017}
A.~Van Pamel, G.~Sha, S.~I.~Rokhlin, and M.~J.~S. Lowe \enquote{{Finite-element modelling of elastic wave propagation and scattering within heterogeneous media},} Proc. R. Soc. \textbf{473}, 2016.0738 (2017).

\bibitem{Rossum1999}
M.~C.~W. van Rossum, and Th.~M. Nieuwenhuizen \enquote{{Multiple scattering of classical waves: microscopy, mesoscopy, and diffusion},} Rev. Mod. Phys. \textbf{71}, 313--371  (1999). 

\bibitem{Meglinski2005}
I.~V.~Meglinski, V.~L. Kuzmin, D.~Y. Churmakov, and D.~A. Greenhalgh \enquote{{Monte Carlo simulation of coherent effects in multiple scattering},} Proc. R. Soc. A.\textbf{461} 43--53 (2005).

\bibitem{RamellaRoman2005}
J.~C. Ramella-Roman, S.~A. Prahl, and S.~L. Jacques, \enquote{{Three Monte Carlo programs of polarized light transport into scattering media: part I},} Opt. Express \textbf{13}, 4420--4438 (2005)

\bibitem{Markel2019}
V.~A. Markel, \enquote{{Extinction, scattering and absorption of electromagnetic waves in the coupled-dipole approximation},} J. Quant. Spectro. Rad. Trans. \textbf{236}, 106611
(2019).


\bibitem{Beenakker1997}
C.~W.~J. Beenakker, \enquote{{Random-matrix theory of quantum transport},} Rev.
  Mod. Phys. \textbf{69}, 731--808 (1997).

\bibitem{Akemann2015}
G.~Akemann, J.~Baik, and P.~{Di Francesco}, eds., \emph{{The Oxford Handbook of
  Random Matrix Theory}} (Oxford University Press, Oxford, 2015).

\bibitem{Aiello2005a}
A.~Aiello and J.~P. Woerdman, \enquote{{Physical Bounds to the
  Entropy-Depolarization Relation in Random Light Scattering},} Phys. Rev.
  Lett. \textbf{94}, 090406 (2005).

\bibitem{Choi2011a}
W.~Choi, A.~P. Mosk, Q.~H. Park, and W.~Choi, \enquote{{Transmission
  eigenchannels in a disordered medium},} Phys. Rev. B \textbf{83}, 1--6
  (2011).

\bibitem{Hanada2018}
M.~Hanada, H.~Shimada, and M.~Tezuka, \enquote{{Universality in chaos: Lyapunov
  spectrum and random matrix theory},} Phys. Rev. E \textbf{97}, 022224 (2018).


\bibitem{Byrnes2020}
N.~Byrnes and M.~R. Foreman, \enquote{{Universal bounds for imaging in
  scattering media},} New J. Phys. \textbf{22}, 083023 (2020).



\bibitem{Garratt2021}
S.~J. Garratt and J.~T. Chalker, \enquote{{Many-Body Delocalization as Symmetry
  Breaking},} Phys. Rev. Lett. \textbf{127}, 026802 (2021).

\bibitem{Aubry2009}
A.~Aubry and A.~Derode, \enquote{{Random matrix theory applied to acoustic
  backscattering and imaging in complex media},} Phys. Rev. Lett. \textbf{102},
  084301 (2009).

\bibitem{Rotter2017}
S.~Rotter and S.~Gigan, \enquote{{Light fields in complex media: Mesoscopic
  scattering meets wave control},} Rev. Mod. Phys.  (2017).

\bibitem{Hsu2017}
C.~W. Hsu, S.~F. Liew, A.~Goetschy, H.~Cao, and A.~{Douglas Stone},
  \enquote{{Correlation-enhanced control of wave focusing in disordered
  media},} Nat. Phys. \textbf{13}, 497--502 (2017).

\bibitem{Wigner1955}
E.~P. Wigner, \enquote{{Characteristic Vectors of Bordered Matrices With
  Infinite Dimensions},} Annal. Math. \textbf{62}, 548 (1955).

\bibitem{Dyson1962}
F.~J. Dyson, \enquote{{Statistical Theory of the Energy Levels of Complex
  Systems. I},} J. Math. Phys. \textbf{3}, 140--156 (1962).

\bibitem{Dyson1962a}
F.~J. Dyson, \enquote{{Statistical Theory of the Energy Levels of Complex
  Systems. II},} J. Math. Phys. \textbf{3}, 157--165 (1962).

\bibitem{Andreev1994}
A.~V. Andreev, B.~D. Simons, and N.~Taniguchi, \enquote{{Supersymmetry applied
  to the spectrum edge of random matrix ensembles},} Nucl. Phys. B
  \textbf{432}, 487--517 (1994).

\bibitem{Byrnes2022a}
N.~Byrnes and M.~R. Foreman, \enquote{{Polarisation statistics of vector
  scattering matrices from the circular orthogonal ensemble},} Opt. Commun.
  \textbf{503}, 127462 (2022).

\bibitem{Goetschy2013}
A.~Goetschy and A.~D. Stone, \enquote{{Filtering random matrices: The effect of
  incomplete channel control in multiple scattering},} Phys. Rev. Lett.
  \textbf{111}, 063901 (2013).

\bibitem{Goetschy2011}
A.~Goetschy and S.~E. Skipetrov, \enquote{{Non-Hermitian Euclidean random
  matrix theory},} Phys. Rev. E \textbf{84}, 011150 (2011).

\bibitem{Goetschy2013a}
A.~Goetschy and S.~E. Skipetrov, \enquote{{Euclidean random matrices and their
  applications in physics},} arXiv: 1303.2880  (2013).

\bibitem{Berk2021b}
J.~Berk and M.~R. Foreman, \enquote{{Role of Multiple Scattering in Single
  Particle Perturbations in Absorbing Random Media},} Phys. Rev. Research
  \textbf{3}, 033111 (2021).

\bibitem{Vivo2008}
P.~Vivo and E.~Vivo, \enquote{{Transmission eigenvalue densities and moments in
  chaotic cavities from random matrix theory},} J. Phys. A: Math. Theor.
  \textbf{41}, 10 (2008).

\bibitem{Mehta2004}
M.~L. Mehta, \emph{{Random Matrices}} (Academic Press, 2004), 3rd ed.


\bibitem{Crisanti} 
A.~Crisanti, G.~Paladin and A.~Vulpiani \emph{Products of Random Matrices in Statistical Physics}, vol 104, Springer Series in Solid-State Sciences (Springer-Verlag, Berlin, 1998)

\bibitem{Tulino2004} A.~Tulino and S.~Verd\'{u} \emph{Random Matrix Theory and Wireless Communications} (now Publishers Inc., Hanover, 2004).

\bibitem{Weidenmuller1990} H.~A.~Weidenm\"{u}ller \enquote{{Scattering theory and conductance fluctuations in mesoscopic systems},} Physica A: Stat. Mech. Appl. \textbf{167}, 28--42 (1990).


\bibitem{Dorokhov1982}
O.~N. Dorokhov, \enquote{{Transmission coefficient and the localization length
  of an electron in N bound disordered chains},} JETP Lett. \textbf{36},
  318--321 (1982).

\bibitem{Mello1988b}
P.~A. Mello, P.~Pereyra, and N.~Kumar, \enquote{{Macroscopic approach to
  multichannel disordered  conductors},} Annal. Phys. \textbf{181}, 290--317
  (1988).

\bibitem{Muttalib2002}
K.~A. Muttalib and V.~A. Gopar, \enquote{{Generalization of the DMPK equation
  beyond quasi one dimension},} Phys. Rev. B \textbf{66}, 115318 (2002).

\bibitem{Douglas2014}
A.~Douglas, P.~Marko{\v{s}}, and K.~A. Muttalib, \enquote{{The generalized DMPK
  equation revisited: Towards a systematic derivation},} J. Phys. A: Math. Gen.
  \textbf{47}, 125103 (2014).

\bibitem{Byrnes2022b}
N.~Byrnes and M.~R. Foreman, \enquote{{Random matrix theory of polarized light
  scattering in disordered media},} Waves Rand. Complex Media, \emph{in press}, 1--29 (2022). https://doi.org/10.1080/17455030.2022.2153305


\bibitem{Efron1979}
B.~Efron, \enquote{{Bootstrap Methods: Another Look at the Jackknife},} Annal. Stat. \textbf{7} 1--26 (1979). 

\bibitem{Efron1982}
B.~Efron, \emph{The Jackknife, the Bootstrap, and Other Resampling Plans} (Philadelphia: SIAM Press, 1982).

\bibitem{Efron1993}
B.~Efron, and R.~J. Tibshirani, \emph{An Introduction to the Bootstrap} (New York: Chapman and Hall, 1993).



\bibitem{Nieuwenhuizen1995}
T.~M. Nieuwenhuizen, and M.~C.~W. van Rossum, \enquote{{Intensity Distributions of Waves Transmitted through a Multiple Scattering Medium},} Phys. Rev. Lett., \textbf{74}, 2674--2677 (1995).

\bibitem{Bicout1992}
D.~Bicout, and C.~Brosseau, \enquote{{Multiply scattered waves through a spatially random medium: entropy production and depolarization},} J. Phys. I France \textbf{2}, 2047--2063 (1992).

\bibitem{Revesz2014}
P.~R\'ev\'esz \emph{The laws of large numbers} (Academic Press, 2014).


\bibitem{Berkovitsa1994}
R.~Berkovits and S.~Feng, \enquote{{Correlations in coherent multiple
  scattering},} Phys. Reports \textbf{238}, 135--172 (1994).

\bibitem{Osnabrugge2017}
G.~Osnabrugge, R.~Horstmeyer, I.~N. Padopoulos, B.~Judkewitz, and I.~M.
  Vellekoop, \enquote{{Generalized optical memory effect},} Optica \textbf{4},
  886--892 (2017).

\bibitem{Byrnes2021b} 
N.~Byrnes and M.~R.~Foreman, \enquote{{Symmetry constraints for vector scattering and transfer matrices containing evanescent components: energy conservation, reciprocity and time reversal},} Phys. Rev. Research \textbf{3}, 013129 (2021).

\bibitem{Fayard2018}
N.~Fayard, A.~Goetschy, R.~Pierrat, and R.~Carminati, \enquote{{Mutual
  Information between Reflected and Transmitted Speckle Images},} Phys. Rev.
  Lett. \textbf{120}, 073901 (2018).

\bibitem{IpsenThesis} J.~R.~Ipsen \emph{Products of Independent Gaussian Random Matrices}, PhD thesis, (Bielefeld University, 2015). 


\bibitem{Bona2012}
M.~Bona, \emph{{Combinatorics of Permutations}} (CRC Press, New York, 2012),
  2nd ed.


\bibitem{ScharfComplexBook} P.~J.~Schreier and L.~L.~Scharf, \emph{{Statistical Signal Processing of Complex-Valued Data - The Theory of Improper and Noncircular Signals}} (Cambridge University Press, Cambridge, 2010.)

\bibitem{oeisA005718}
The On-Line Encyclopedia of Integer Sequences, Sequence A005718 \url{https://oeis.org/A005718}.

\bibitem{oeisA107991}
The On-Line Encyclopedia of Integer Sequences, Sequence A107991 \url{https://oeis.org/A107991}.


\bibitem{Leonov1959}
V.~P. Leonov and A.~N. Shiryaev, \enquote{{On a Method of Calculation of Semi-Invariants},} Theo. Prob. Appl. \textbf{4}, 319--329 (1959).

\bibitem{Eynard1998}  B.~Eynard and M.~L.~Mehta, \enquote{{Matrices coupled in a chain: {I}. Eigenvalue correlations},} J. Phys. A: Math. Gen. \textbf{31}, 4449-4456 (1998).

\bibitem{Mahoux1998} 
 G.~Mahoux, M.~L.~Mehta and J.-M.~Normand \enquote{{Matrices coupled in a chain: {II}. Spacing functions},} J. Phys. A: Math. Gen., \textbf{31}, 4457--4464 (1998).

\bibitem{Akeman2013} G.~Akemann, M.~Kieburg and L.~Wei, \enquote{{Singular value correlation functions for products of {W}ishart random matrices},} J. Phys. A: Math. Theor. \textbf{46} 275205 (2013).

\bibitem{Mehta1994} 
M.~L.~Mehta and P.~Shukla, \enquote{{Two coupled matrices: eigenvalue correlations and spacing functions},} J. Phys. A: Math. Gen. \textbf{27}, 7793--7803 (1994). 

\bibitem{Furstenberg1960} 
H.~Furstenberg and H.~Kesten, \enquote{{Products of Random Matrices},}
Annal. Math. Stat. \textbf{31}, 457--469 (1960). 

\bibitem{Akeman2023}
G. Akemann, and S.~S.~Byun \enquote{{The Product of $m$ Real $N\times N$
Ginibre Matrices: Real Eigenvalues in the Critical Regime, $m=O(N)$},} Constr. Approx. (2023). https://doi.org/10.1007/s00365-023-09628-2


\bibitem{supp} See Supplemental Material at [URL will be inserted by publisher] for step by step animations of proof for each Class.

\bibitem{Stanley}
R.P.~Stanley,
\enquote{{Two enumerative results on cycles of permutations},}
European J. Combin.,
\textbf{32}, 937--943
(2011).

\bibitem{Zagier}
D.~Zagier,
\enquote{{On the distribution of the number of cycles of elements in symmetric groups},}
Nieuw Arch. Wisk., \textbf{13}, 489--495 (1995).



\end{thebibliography}
\end{document}